\documentclass[usenatbib]{mnras}

\usepackage{setspace}
\usepackage{siunitx}
\usepackage{here}
\usepackage[T1]{fontenc}
\usepackage{ae,aecompl}
\usepackage{enumerate}

\usepackage{graphicx}	
\usepackage{amssymb}	
\usepackage{multirow}
\usepackage{comment}
\usepackage{url}
\usepackage{threeparttable}
\usepackage[hang,small]{caption}
\usepackage{newtxtext,newtxmath}
\usepackage{color}
\title[Cyg~A east hot spot with \textit{Herschel}]{\textit{Herschel} discovery of far-infrared emission from the hot spot D in the radio galaxy Cygnus~A}
\author[Y. Sunada et al.]{
Yuji Sunada,$^{1}$\thanks{E-mail: sunada@heal.phy.saitama-u.ac.jp}
Naoki Isobe,$^{2}$ 
Makoto S. Tashiro,$^{1,2}$ 
Motoki Kino,$^{3,4}$ 
Shoko Koyama,$^{5,6}$
\newauthor
 and Satomi Nakahara$^{2}$
\\
$^{1}$ Depertment of Physics, Saitama University, 255 Shimo-Okubo, Sakura-ku, Saitama, 338-8570, Japan\\
$^{2}$ Institute of Space and Astronautical Science (ISAS), Japan Aerospace Exploration Agency (JAXA),\\
3-1-1 Yoshinodai, Chuo-ku, Sagamihara, Kanagawa, 252-5210, Japan\\
$^{3}$ Kogakuin University of Technology \& Engineering, Academic Support Center, 2665-1 Nakano, Hachioji, Tokyo 192-0015, Japan\\
$^{4}$ National Astronomical Observatory of Japan, 2-21-1 Osawa, Mitaka, Tokyo, 181-8588, Japan\\
$^{5}$ Niigata University, 8050 Ikarashi-nino-cho, Nishi-ku, Niigata 950-2181, Japan\\
$^{6}$ Institute of Astronomy and Astrophysics, Academia Sinica, \\
11F of Astronomy-Mathematics Building, AS/NTU No. 1, Sec. 4, Roosevelt Rd, Taipei 10617, Taiwan, R.O.C.\\
}

\date{Accepted XXX. Received YYY; in original form ZZZ}

\pubyear{}

\begin{document}
\label{firstpage}
\pagerange{\pageref{firstpage}--\pageref{lastpage}}
\maketitle


\begin{abstract}
The far infrared counterpart of hot spot D, the terminal hot spot of the eastern jet hosted by the radio galaxy Cygnus A, is detected with \textit{Herschel}
Aperture photometery of the source performed in 5 photometric bands covering the wavelength range of $70$--$350$ \micron.
After removing the contamination from another nearby hot spot, E, the far-infrared intensity of hot spot D is derived as $83\pm13$ and $269\pm66$ mJy at $160$ and $350$ \micron , respectively. 
Since the far-infrared spectrum of the object smoothly connects to the radio one, the far-infrared emission is attributed to the synchrotron radiation from the radio-emitting electron population. 
The radio-to-near-infrared spectrum is confirmed to exhibit a far-infrared break feature at the frequency of $\nu_\mathrm{br}=2.0^{+1.2}_{-0.8} \times10^{12}$ Hz. 
The change in energy index at the break ($\Delta\alpha=0.5$) is interpreted as the impact of radiative cooling on an electron distribution sustained by continuous injection from diffusive shock acceleration.
By ascribing the derived break to this cooling break, the magnetic field, $B$, in the hot spot is determined as a function of  its radius, $R$ within a uniform one-zone model combined with the strong relativistic shock condition. 
An independent $B$-$R$ constraint is obtained by assuming the X-ray spectrum is wholly due to synchrotron-self-Compton emission.
By combining these conditions, the two parameters are tightly determined as $B=120$--$150$ $\mathrm{\mu G}$ and $R=1.3$--$1.6$ kpc. 
A further investigation into the two conditions indicates the observed X-ray flux is highly dominated by the synchrotron-self-Compton emission.
\end{abstract}


\section{Introduction}\label{sec:int}
Hot spots of Fanaroff-Riley type-II \citep[FR-II;][]{fanaroff74}
radio galaxies are compact and bright radio structures located 
where their jets terminate \citep{meisenheimer89}.
It is widely thought that particles are efficiently accelerated 
into the relativistic regime in the hot spots
via the diffusive shock acceleration (or the so-called Fermi-I process e.g., \citealt{bell78}),
as observationally evidenced by their synchrotron radiation \citep{meisenheimer89}.
Thus, the hot spots are frequently proposed 
as one of the possible production sites of ultra-high-energy cosmic rays
\citep{hillas84,kotera11}.
For detailed investigation into the particle acceleration phenomena 
in the hot spots, 
it is important to measure their magnetic field strength 
since it is theoretically predicted to be one of the major parameters 
which control the particle acceleration and associated cooling efficiencies. 

Since pioneering discoveries of X-ray emission 
from a number of hot spots with \textit{ROSAT} and \textit{Chandra} 
\citep[][]{harris94,wilson00,harris00,hardcastle01,hardcastle02},
a comparison between the radio and X-ray fluxes has been the standard method
to evaluate their magnetic field for more than two decades.
The radio and X-ray spectra of these hot spots are widely 
interpreted by the one-zone Synchrostron-Self-Compton (SSC; \citealt{band85}) process \citep[e.g.,][]{hardcastle04,kataoka05},
except for a few optical hot spots \citep[e.g,][]{wilson01,hardcastle04}.
These studies indicate a typical magnetic field of 
$B=100$--$500$ $\mathrm{\mu G}$ in the hot spots,
which implies nearly equipartition value 
as $B/B_\mathrm{eq} \sim0.1$--$2$.
In addition, the SSC analysis, 
in combination with detailed investigation of the shock dynamics, 
was theoretically applied to infer the energetics and plasma composition 
in the hot spots,
and hence, in the jets of the FR-II radio galaxies \citep[e.g.,][]{kino04,godfrey13,snios18,croston18,sikora20}.

It is not necessarily well recognised that 
the standard method is often accompanied by a "hidden" uncertainty. 
The synchrotron and SSC intensities are mainly dependent 
on the following three physical quantities;
i.e., the magnetic field, electron energy density and size of the emission region \citep{band85}.
From measurements of the radio and X-ray fluxes alone,
it is not possible to disentangle these three parameters. 
Thus, the observed radio and X-ray spectra are inevitably reproduced 
by the SSC model with different parameter sets 
\citep{wilson00,kino04,stawarz07}.
In order to overcome this difficulty, the apparent source size is usually estimated 
by making use of high-resolution images.
However, for some hot spots, a difference by more than a factor of two is found 
among measurements on the source size 
due to the different observational configurations \citep[e.g.,][]{perley84,wilson00}.
This results in a notable uncertainty in the magnetic field estimation.
Therefore, we need to search for another observational indicator 
of the magnetic field and/or the source size of the hot spots.

We propose to adopt a synchrotron spectral feature, called the "cooling break",
to be observed in the frequency range higher than the radio one 
as the third item of observational information 
to constrain the magnetic field and size of the hot spots.
In the case of the diffusive shock acceleration 
under a continuous energy injection condition,
the spectrum of the accelerated electrons is predicted 
to exhibit a break at the Lorentz factor 
for which the radiative cooling time scale is balanced 
with the adiabatic loss one \citep{meisenheimer89,carilli91}.
As a result, the synchrotron spectrum is expected to break similarly 
at the corresponding frequency.
This cooling break is utilised to estimate the magnetic field and source size
\citep{inoue96,kino04}.
In fact, the cooling break is successfully adopted to model the multi-frequency spectra of blazars \citep[e.g.][]{kataoka00}.
Because the hot spots and blazars are widely considered to share injection and cooling processes,
the hot spots are expected to exhibit inevitably the cooling break in their synchrotron spectrum.

From synchrotron studies of several archetypical hot spots in the near-infrared (NIR) to the optical bands  \citep{meisenheimer97,kraft07,werner12},
their cooling break is suggested to be located in the far-infrared (FIR) band, 
namely $10^{11}\-- 10^{12}$ Hz \citep{brunetti03,cheung05}.
However, their radio-to-optical spectrum is alternatively
interpreted by the spectral cut-off feature corresponding to the maximum electron energy, 
rather than the break one, so far.
A definite observational evidence of the cooling break in the spectrum of the hot spots 
has remained unexplored,
because a wide spectral gap around the FIR band in the spectrum of the hot spots
prevents from discriminating between the two spectral features. 

In order to search for the cooling break of the hot spots,
the Spectral and Photometric Imaging REceiver 
\citep[SPIRE;][]{griffin10} and Photodetector Array Camera and Spectrometer  \citep[PACS;][]{poglitsch10} onboard the \textit{Herschel} space observatory \citep{pilbratt10} provides an ideal tool.
A combination of these two instruments operated in the photometer mode 
covers a wide FIR wavelength range of $70$--$500$ $\mathrm{\mu m}$, 
corresponding to the frequency range of $(0.6$--$4.3) \times 10^{12}$ Hz, 
where the hot spots are typically expected to exhibit the cooling break as mentioned above.
Thanks to their reasonable photometrical sensitivity (typically $\gtrsim 10$ mJy) and 
moderate angular resolution (about 20 and 10 arcsec for SPIRE and PACS, respectively),
these two instruments are applicable to the hot spots 
with a projected distance from its nucleus of larger than a few arcmin.
In fact, the FIR emission from the western hot spot of the representative FR-II radio galaxy 
Pictor A was discovered with the \textit{Herschel} SPIRE \citep{isobe20}.

The hot spot named "D" of the FR-II radio galaxy Cygnus A \citep{carilli96}, located at the end of its eastern jet 
\citep{bentley75,perley84} is one of the best candidates for the FIR studies with \textit{Herschel}.
This hot spot has been extensively observed in radio, NIR and X-ray bands \citep{meisenheimer89,carilli91,wilson00,Wright04,stawarz07}.
Its radio-to-X-ray spectrum is usually explained by the SSC process \citep{harris94,wilson00,stawarz07}.
By artificially introducing a possible cooling break at the frequency of 0.5$\times$10$^{12}$ Hz
into the SSC model,
\cite{stawarz07} estimated the magnetic field strength of the hot spot 
as $B=270$ $\mathrm{\mu G}$ for the radius of $R=0.8$ kpc.
The observed properties of hot spot D 
are utilised to investigate the properties of the jets, lobes and cocoons 
of this radio galaxy \citep[e.g.,][]{wilson06,yaji10}.
Analytical and theoretical studies tried to evaluate 
the dynamics, energetics and plasma composition related to the jets, 
by connecting the physical quantities observed from the above components
\citep[e.g.,][]{smith02,kino05,ito08,kino12,steenbrugge14,kawakatu16,snios18}.
However, as mentioned above, due to the FIR spectral gap in $10^{11}$--$10^{13}$ Hz, 
the cooling break has not yet been observationally confirmed, 
and there probably remain notable uncertainties in the  physical parameters 
of the hot spot.
From its relatively high radio flux density of $\sim10$ Jy at 10 GHz and 
radio energy index of $\alpha\sim1$,
the FIR flux density of the hot spot is estimated as $\sim 100$ mJy at $10^{12}$ Hz,
which is significantly higher than the SPIRE and PACS sensitivities.
Because of its large angular separation from the nucleus of Cygnus A, of $\sim50$ arcsec, 
hot spot D is safely detectable by the two instruments 
without suffering from any nuclear contamination.
Therefore, we investigate the FIR infrared spectrum of the hot spot to make sure 
of its cooling break, by utilising the \textit{Herschel} SPIRE and PACS data. 

For direct comparison,
we adopt the same cosmological constants as those in \citet{stawarz07}
throughout the present paper;
i.e., $H_0=71$ km s$^{-1}$ Mpc$^{-1}$, $\Omega_\mathrm{m}=0.27$, $\Omega_\Lambda=0.73$.
At the redshift  of Cygnus A \citep[$z=0.05607$;][]{owen97},
the luminosity distance is evaluated as $248$ Mpc.
The angular size of 1 arcsec corresponds to the physical size of 1.07 kpc 
at the source rest frame.

\section{Observation and Analysis}
\label{sec:obs_ana}
\subsection{\textit{Herschel} Data}\label{sec:data}
In order to investigate the FIR properties of the hot spot of Cygnus A, we utilised the FIR data obtained with the PACS and SPIRE photometers.
The SPIRE photometer mapped Cygnus A on 2011 October 11 in the Large Map mode(Obs. ID of 1342230853), while the PACS photometers observed the object on 2011 December 24 with two pairs of cross-scan mappings, at a scan speed of 20 arcsec s$^{-1}$.
In the first cross-scan observation (Obs. ID of 1342235110 and 1342235111), the blue camera of the PACS photometer was operated at the wavelength of 70 \micron, while in the second one (Obs. ID of 1342235112 and 1342235113) the  100 \micron \ filter was adopted for the blue camera.

We retrieved the final release of the SPIRE and PACS science products from the Herschel Science Archive.
We analysed the data with version 15.0.1 of the {\small HERSCHEL INTERACTIVE PROCESSING ENVIRONMENT}, by applying the corresponding SPIRE and PACS calibration trees, {\small SPIRE\_CAL\_14\_3} and {\small PACS\_CAL\_77\_0}, respectively.
We adopted the Level-2 science products for SPIRE imaging and photometry.
For the PACS blue camera (i.e., 70 and 100 \micron), we employed the Level-2.5 data, while we utilised the Level-3 products for the PACS red camera (160 \micron).
From the PACS image products, we analysed the HPPUNIMAP and HPPJSMAP maps, both of which are applicable to extended sources.
We confirmed that the photometric results from the two maps were consistent with each other within 5 per cent.
We, hereafter, adopt the HPPUNIMSP results because its photometric accuracy is slightly better at 70 \micron.

\subsection{FIR images}
\label{sec:image_ana}
\begin{figure*}
    \includegraphics[width=\linewidth]{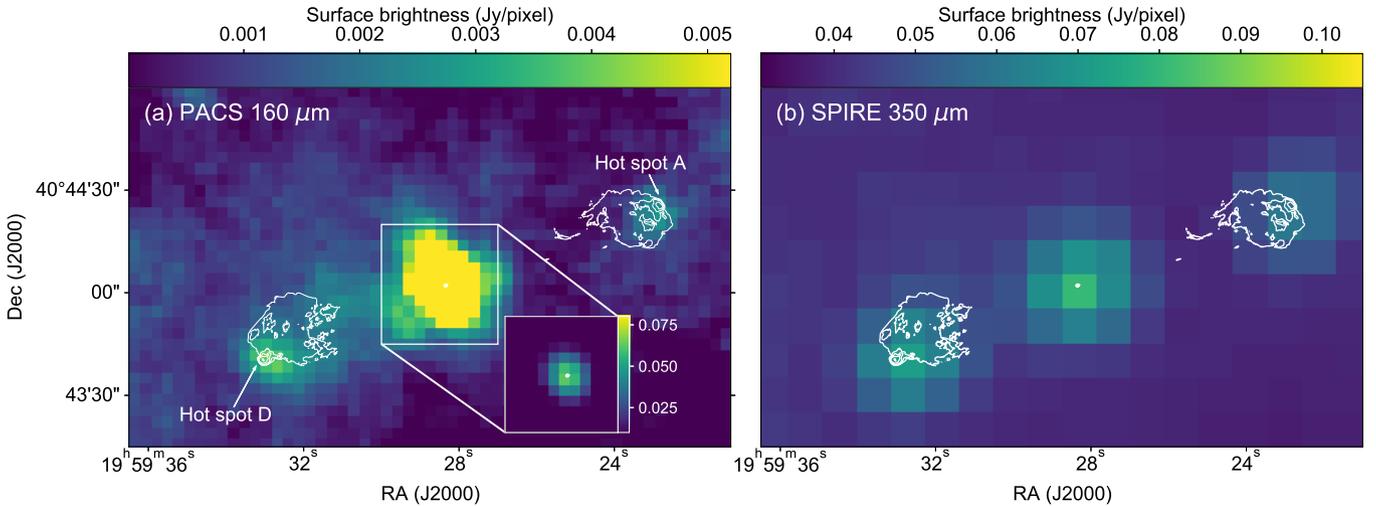} 
    \caption{Panel (a) shows the 160 \micron \ PACS image of Cygnus A. Panel (b) shows the 350 \micron \ SPIRE image. On both panels, the 5 GHz radio contours \citep{perley84} are overlaid. The coloured bar above each panel shows the surface brightness in the unit of Jy pixel$^{-1}$. The arrows in panel (a) show the hot spots A and D. The inset in panel (a) shows the PACS image in a different colour scale and VLA contours around the nucleus. \label{fig:image}}
\end{figure*}

Figure \ref{fig:image} shows the 160 \micron \ PACS (panel a) and 350 \micron \  SPIRE (panel b) images around the radio galaxy Cygnus A on which the 5 GHz radio contours \citep{perley84} are overlaid.
The FIR emission from the nucleus of Cygnus A is significantly detected both in the PACS and SPIRE bands.
The nucleus in the PACS image is saturated for clear visualisation of fainter emission associated with the hot spots D.
The inset inserted into panel (a) of Figure \ref{fig:image} shows the image around the nucleus with a different colour scale.
The PACS image was aligned with the radio one by referring to the nucleus position to correct the astrometric error ($\lesssim$ 1 arcsec for all the PACS bands).
We ignored the astrometric error for all the SPIRE bands, because the typical astrometric uncertainty of \textit{Herschel} (2 arcsec; \citealt{swinyard10}) is significantly smaller than the SPIRE pixel size (6--14 arcsec, dependent on the photometric band).
The FIR sources associated with the radio hot spots A and D were cleary detected.
In this paper, we focus on the FIR source corresponding to hot spot D because the hot spot A is reported to be subjected to bright emission surrounding it in the NIR and optical bands \citep{stawarz07}.

\begin{figure}
  \begin{center}
    \includegraphics[width=\linewidth]{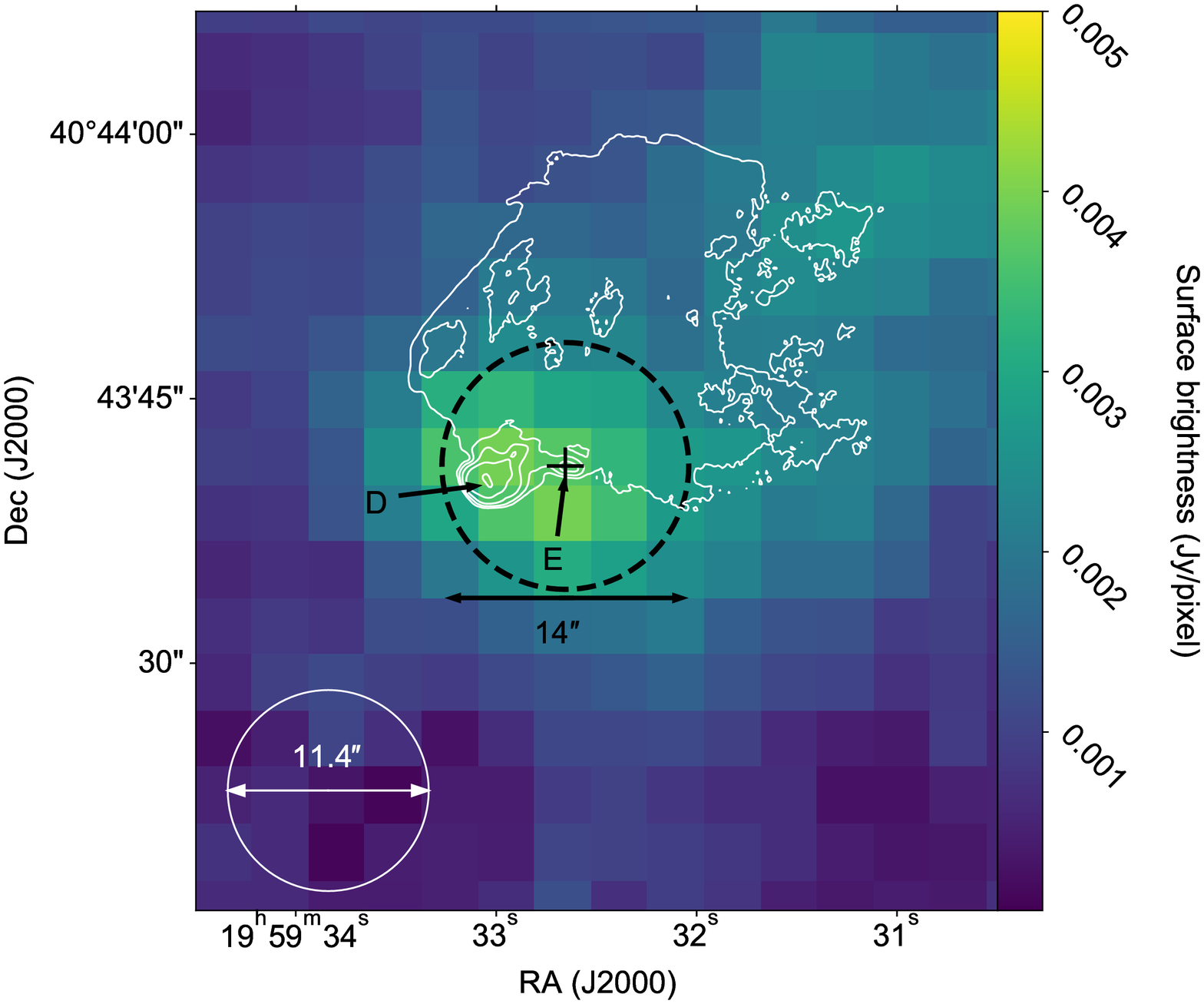} 
    \caption{The 160 $\mathrm{\mu m}$ image around the FIR source associated to hot spot D observed by the PACS. The black dashed circle and the black cross indicate the apparent source size and position obtained by SUSSExtractor, respectively.
    The black arrow below the dashed circle indicates the source size of 14 arcsec.
    The white solid circle shows the PSF size (11.4 arcsec) in the FWHM of PACS at 160 \micron . The arrows labeled as D and E indicate the position of radio hot spots D and E, respectively.
    \label{fig:loc}}
  \end{center}
\end{figure}

Firstly, we investigated the spatial property of the FIR source using the PACS image, which has a better angular resolution than that of SPIRE.
Figure \ref{fig:loc} shows the 160 \micron \ PACS close-up view around hot spot D.
The FIR source seems slightly extended in comparison to the PACS point spread function (PSF) with the size of $d_\mathrm{PSF}$=11.4 arcsec in the Full With at Half Maximum (FWHM) at 160 \micron, shown with the white circle in Figure \ref{fig:loc}.
The radio image reveals the fainter hot spot E, $\sim$ 5 arcsec to the west of hot spot D.
Thus, in the PACS image, the FIR source is possibly contaminated by hot spot E.

In order to roughly evaluate of the size of the FIR source, 
we firstly applied SUSSEXtractor \citep{savage07} to the PACS data,
by changing the apparent source size.
No FIR source was found when the input source size was set to the PSF size.
However, we detected the source for an input source size 
larger than the PSF size, 
except for a marginal detection at 70 \micron.
Table \ref{tab:pos} summarises the PACS result of the source detection.
Here, we calculated the apparent source diameter, $d_\mathrm{ap}$,
which gives the highest Signal-to-Noise (SN) ratio.
The obtained source size ($d_\mathrm{ap} = 13$ arcsec at 160 \micron),
which is larger than the PSF one ($d_\mathrm{PSF} = 11.4$ arcsec in the FWHM), 
implies the possible source extension.

To measure the apparent source size more quantitatively,  
we fitted an axially symmetric two-dimensional Gaussian function 
with a flat offset to the PACS image of the source.
Here, we analysed only the 160 \micron\ image,
since it gives the highest SN ratio among the three PACS bands.
The central position of the Gaussian function is fixed 
at the source position derived by SUSSEXtractor.
The FWHM apparent source size is evaluated 
as $d_\mathrm{fit}=14\pm1$ arcsec.
It is confirmed that the apparent source size is larger than the PSF size.
We deconvolved the PSF size from the obtained appernt source size 
as $d_\mathrm{src} = \sqrt{d_\mathrm{fit}^2-d_\mathrm{PSF}^2}$.
After the deconvolution,
the derived source size ($d_\mathrm{src}$ = $8\pm2$ arcsec) 
becomes comparable to the angular separation 
between hot spots D and E.

The FIR source position determined by SUSSEXtractor is significantly shifted from radio hot spot D, 
and the angular offset of ($\Delta\theta_\mathrm{RA}$, $\Delta\theta_\mathrm{Dec}$) = ($-4.3\pm1.1$, $0.8 \pm 1.1$) arcsec at 160 \micron \ 
 is not negligible in comparison with the PSF size.
In Figure \ref{fig:loc}, we show the FIR source position determined with SUSSEXtractor 
and the best-fit apparent source size (i.e., $d_\mathrm{fit}$)
with the black cross and dashed black circle, respectively.
Although hot spot D is on the brightest pixel of the PACS image, 
the source is shifted toward radio hot spot E.
Besides, the apparent source size encompasses both hot spots D and E.
These imply the FIR emission mainly originates in hot spot D with a notable contamination from hot spot E.
Both the deconvolved source size and the position offset implies some contamination from hot spot E.

The FIR source seems point-like in the SPIRE image displayed in panel (b) of Figure \ref{fig:image}.
The point-like nature of the FIR source in the SPIRE band seems reasonable, since the  deconvolved source size with the PACS (i.e., $8\pm2$ arcsec at 160 \micron) is smaller than the SPIRE PSF (see Table \ref{tab:pos}).
Thus, we performed SUSSEXtractor by adopting the SPIRE PSF size.
We significantly detected the source with an SN ratio is 70 in the 350 \micron \ image.
Table \ref{tab:pos} also summarises the result of the SPIRE source detection.
The derived angular offset of ($\Delta\theta_\mathrm{RA}$, $\Delta\theta_\mathrm{Dec}$) = ($-1.3\pm0.3$, $1.6 \pm 0.3$) arcsec at 250 \micron \ is negligible in comparison with the PSF size of $d_\mathrm{PSF}$ = 24 arcsec even at 250 \micron, \ where the spatial resolution is highest among the SPIRE photometric bands.
Thus, we did not resolve hot spots D and E due to the SPIRE angular resolution.

\begin{table*}
    \begin{center}
    \caption{Summary of the FIR source spatial properties\label{tab:pos} 
    derived with SUSSEXtractor}
    \begin{tabular}{lccccccc}
    \hline
    \multirow{2}{*}{instrument} & $\mathrm{\lambda^{\ a}}$ &${d_\mathrm{PSF}}^\mathrm{b}$& ${\Delta\theta_\mathrm{RA}}^\mathrm{c}$& ${\Delta\theta_\mathrm{Dec}}^\mathrm{c}$
    &${d_\mathrm{ap}}^\mathrm{d}$ & \multirow{2}{*}{SN ratio}\\
    &(\micron) &FWHM (arcsec)& arcsec & arcsec &FWHM (arcsec) &\\
    \hline
    SPIRE & 500 & 35 & $-1.5 \pm 0.4$ & $3.5 \pm 0.4$  & 35 (fixed) & 56 \\
          & 350 & 24 & $-1.3 \pm 0.3$ & $1.6 \pm 0.3$  & 24 (fixed) & 70 \\    
          & 250 & 18 & $-2.7 \pm 0.4$ & $1.0 \pm 0.4$  & 18 (fixed) & 29 \vspace{1pt} \\    
     PACS & 160 & 11.4& $-4.3 \pm 1.1$ & $0.8 \pm 1.1$ & 13 & 7.4 \vspace{1pt}\\
          & 100 & 6.8 & $-2.5 \pm 1.1$ & $0.2 \pm 1.1$ & 12 & 6.7 \vspace{1pt}\\
          &  70 & 5.6 & $-2.5 \pm 1.2$ & $0.4 \pm 1.2$ & 9 & 4.5  \vspace{1pt}\\
    \hline
    \multicolumn{7}{l}{a The effective wavelength}\\
    \multicolumn{7}{l}{b The averaged PSF size,, taken from the calibration tree}\\
    \multicolumn{7}{l}{c The angular offsets of the FIR source from the brightness centre of hot spot D at 5 GHz.}\\
    \multicolumn{7}{l}{d The apparent source size, 
    for which the SN ratio of the source is maximised.}\\
    \end{tabular}
    \end{center}
\end{table*}

\subsection{FIR photometry}
\label{se:phot}
We measure the PACS and SPIRE flux of the FIR source associated with hot spot D.
Since hot spot E is not fully resolved with the PACS and SPIRE, as shown in section \ref{sec:image_ana}, we here evaluate the sum flux of hot spots D and E by adopting aperture photometry.
We decompose the source flux into those of D and E by utilising the multi-wavelength spectra in section \ref{sec:spec_D}.

\begin{figure}
    \includegraphics[width=\linewidth]{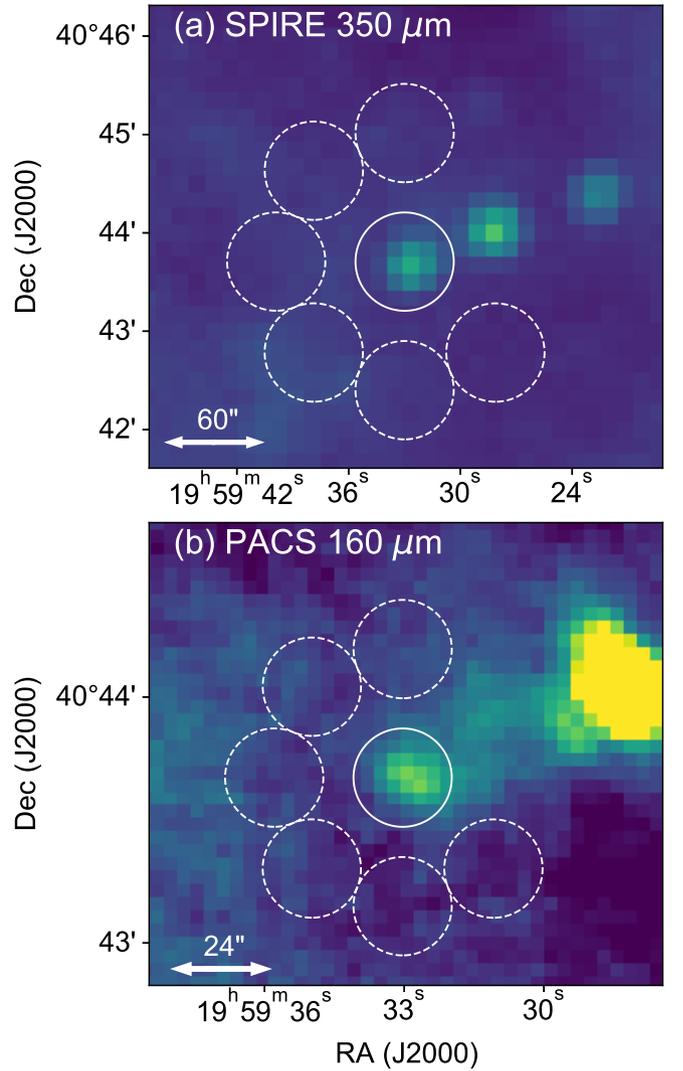} 
    \caption{The source and background apertures used for the photomery with SPIRE (panel a) and PACS (panel b). The solid circle indicates the source aperture with a radius of 30 and 12 arcsec in 350 and 160 \micron \ images, respectively. The dashed circles correspond to the background apertures with the same radius as the source one. The white arrows indicate the aperture diameter. \label{fig:region}}
\end{figure}

The source and background apertures we adopted for the PACS photometry at 160 \micron \ is shown in panel (a) of Figure \ref{fig:region}, while those for the SPIRE photometry at 350 \micron \ is indicated in panel (b).
The radius of the source aperture for the individual SPIRE and PACS photometric bands is summarised in Table \ref{tab:int}.
For the SPIRE photometry, the FIR source position obtained from SUSSEXtractor is employed as the centre of the source aperture.
The radius of the source aperture is determined to avoid contamination from the nucleus.
To evaluate the spatial fluctuation of the background flux density, the six background apertures are selected from the region around the FIR source.
We did not put any background aperture between the FIR source and nucleus to avoid contamination from the nucleus and possibly from the lobe.
For the background regions, we adopted the same aperture radius as for the source one.
At the 500 \micron \ SPIRE band, we did not evaluate the FIR source flux, because the nuclear contamination is expected to be severe due to the poor angular resolution.
For the PACS photometry, the radio position of hot spot D is employed as the source position, instead of that determined with SUSSEXtractor because it nearly coincides with the brightness peak of the FIR source on the PACS image.
The source aperture size is determined to safely include the emission from hot spots D and E but excludes the nuclear contamination and diffuse emission possibly associated with the lobe.
Similar to the SPIRE photometry, the same radius as for the source region is adopted for the background regions.

We performed the SPIRE and PACS photometry by using the individual background regions.
Following the standard manner, we adopted the standard deviation of the fluxes of the six background regions as the photometric error.
The aperture correction was performed to the measured flux densities by retrieving the correction factor for the adopted source radius from the calibration tree.
We adopted the power-law (PL) spectrum with an energy index of $\alpha$ = 1 for the colour correction.
At least in the range of $\alpha$ = 1--2, the colour correction has only a minor contribution with a correction factor of < 1 and < 0.1 per cent for SPIRE and PACS, respectively.
In Table \ref{tab:int}, we summarise the aperture-and-colour corrected flux density ($F_\nu$) of the FIR source in Table 2, together with the adopted correction factor ($C_\mathrm{cor}$).
The corrected flux densities are derived as $F_\nu=$199$\pm$58 and 92.3$\pm$13 mJy 
in the SPIRE 350 \micron \ and the PACS 160 \micron \ photometric bands, respectively. 

\begin{figure}
  \begin{center}
        \includegraphics[width=\linewidth]{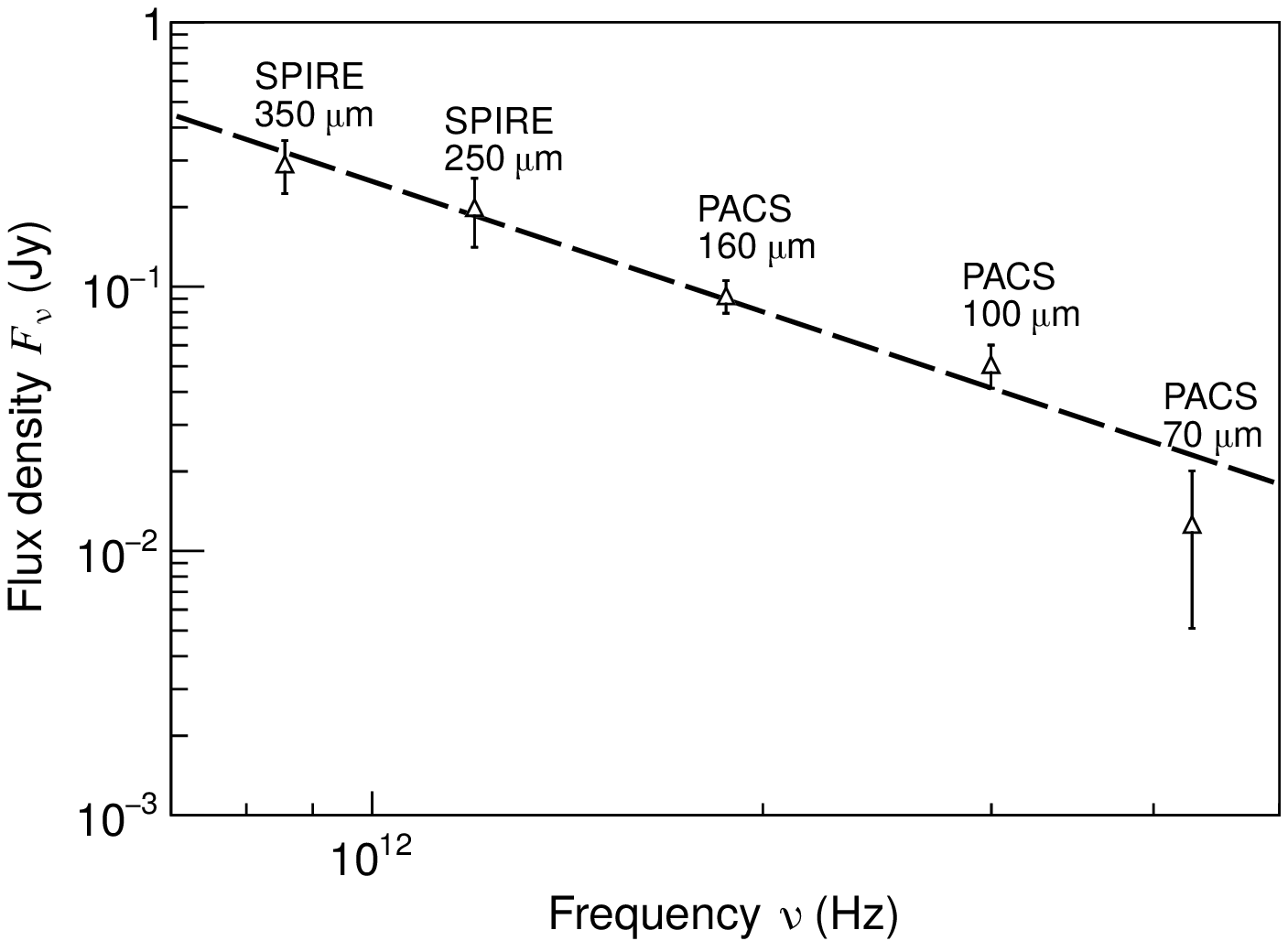} 
    \caption{The FIR spectrum of the source associated with hot spot D obtained in the PACS and SPIRE phtometry, before subtracting the contamination from hot spot E. The open triangles show the flux density of the source, while the dashed line presents the best-fit PL model, with the best-fit energy index of $\alpha=1.64\pm0.17$.\label{fig:hspec}}
  \end{center}
\end{figure}

We show the obtained FIR spectrum of the source in Figure \ref{fig:hspec}.
The FIR spectrum is successfully reproduced by the PL model ($\chi^2$/d.o.f. = 3.55/3), as shown with the dashed line in Figure \ref{fig:hspec}.
An energy index of $\alpha=1.64\pm0.17$ and a flux density at $10^{12}$ Hz of $0.26\pm0.04$ Jy are obtained.
Although the measured energy index is higher than that adopted for the colour correction ($\alpha$ = 1), the difference in the correction factor between these indices is negligible.

\begin{table}
  \begin{center}
  \caption{Summary of the SPIRE and PACS photometry\label{tab:int}}
    \begin{tabular}{lccccc}\hline
        \multirow{2}{*}{instrument} & $\mathrm{\lambda}$ & $\nu^{\mathrm{a}}$ & aperture radius &  \multirow{2}{*}{${C_\mathrm{cor}}^\mathrm{b}$}& ${F_\nu}^{\mathrm{c}}$\\
      &$\mathrm{\mu m}$ & $10^{11}$Hz & arcsec & & mJy \\
      \hline
        SPIRE
        & 350 & 8.6 & 30 & 1.240 & 291 $\pm$ 66\\
        & 250 & 12 & 22 & 1.283 & 199 $\pm$ 58\\
        PACS
        & 160 & 19 & 12 & 1.49  & 92.3 $\pm$ 13\\
        & 100 & 30 & 12 & 1.29 & 50.7 $\pm$ 9.4\\
        &  70 & 43 & 12 & 1.25 & 12.6 $\pm$ 7.5\\
       \hline
       \multicolumn{6}{l}{a The frequency corresponding to the effective wavelength of $\lambda$}\\
       \multicolumn{6}{l}{b The correction factor including the colour and aperture correction}\\
       \multicolumn{6}{l}{c The aperture and colour corrected flux density of the FIR source}\\
    \end{tabular}
  \end{center}
\end{table}

\section{Spectrum of hot spot D}
\subsection{Subtraction of the contamination from hot spot E}
\label{sec:spec_D}
\begin{figure}
  \begin{center}
    \includegraphics[width=\linewidth]{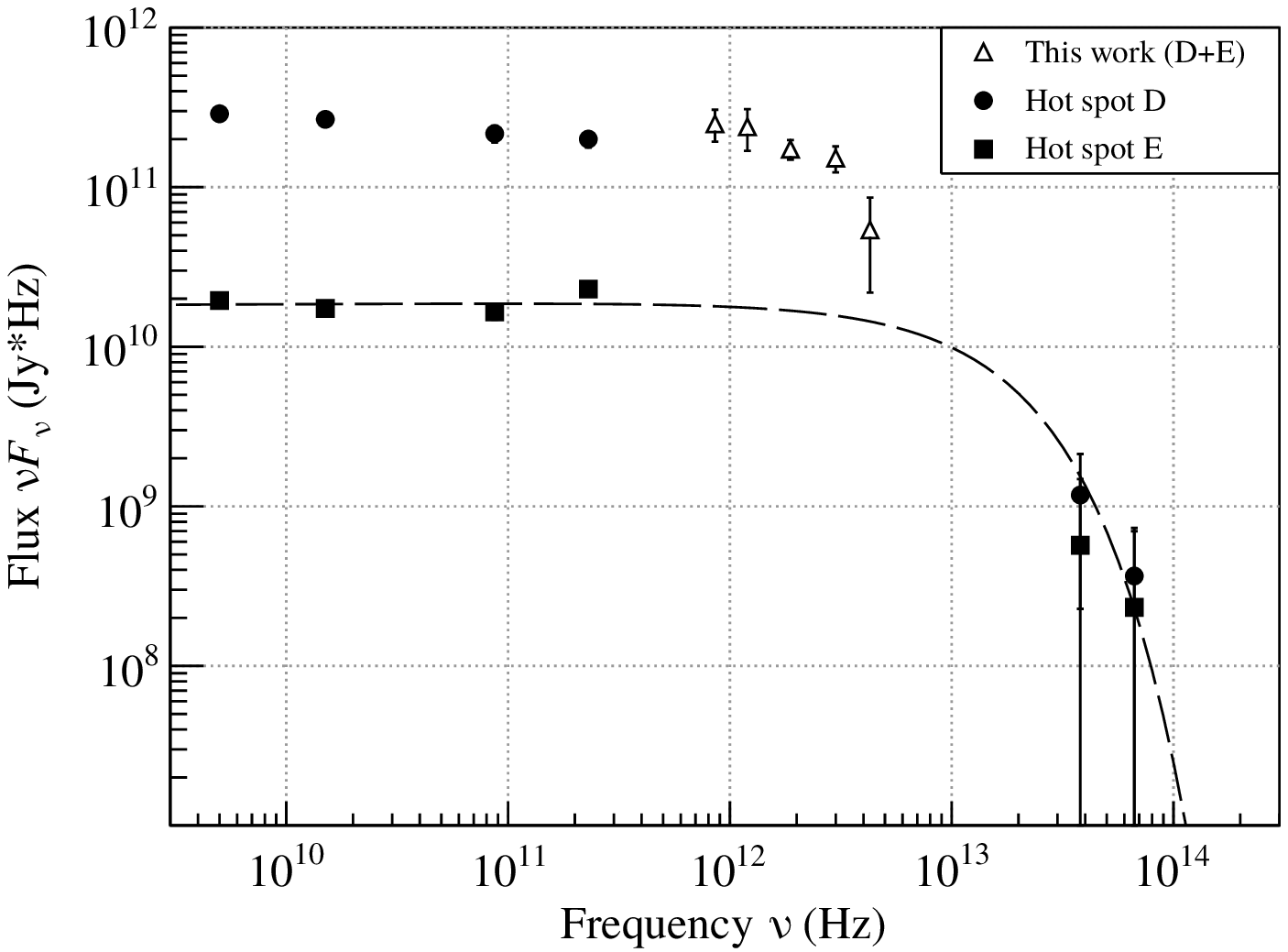}
    \caption{Broadband SED of hot spot D and E. The open trangles show the SED of the FIR source associated with hot spot D with the contamination from the hot spots included, obtained in this work. The filled circles and filled squares show the spectra of hot spots D and E \citep{lazio06,Wright04,stawarz07}, respectively, in the radio and NIR bands. The dashed line displays the best-fit CPL model to the spectrum of hot spot E.\label{fig:flux_all}}
  \end{center}
\end{figure}

In order to investigate the spectral properties of the FIR source, we compare its FIR Spectral Energy Distribution (SED) with the radio and NIR data of hot spot D and E, taken from \cite{lazio06}, \cite{carilli91}, \cite{Wright04} and \cite{stawarz07}.
We show the radio to NIR SED of hot spots D and E in Figure \ref{fig:flux_all}.
In the radio band, both hot spots D and E exhibit a flat SED with an energy index of $\alpha \sim 1$, while of hot spot D is brighter than that of hot spot E by an order of magnitude.
Except for the 70 \micron\ band, \ the obtained FIR flux is slightly higher than  a simple PL extrapolation of the radio flux from hot spot D.
These results suggests that the FIR flux of the source is dominated by the emission from hot spot D with a possible contamination from hot spot E.

To subtract the contamination from hot spot E, we estimate the FIR spectrum of hot spot E by interpolating the radio and NIR spectrum.
hot spot E has a lower NIR flux than the simple PL extrapolation of the radio spectrum.
Thus, we fitted the radio to NIR spectrum of hot spot E with the cut-off power law (CPL) model described as $F_\nu \propto \nu^{-\alpha} \exp{(-\nu / \nu_\mathrm{c})}$ , where $\nu_c$ denotes the cut-off frequency.
The dashed line in Figure \ref{fig:flux_all} shows the best-fit CPL model ($\chi^2/\mathrm{d.o.f.}=6/3$),
with the flux density at 10 GHz of $F_\nu$(10 GHz) = $1.8 \pm 0.1$ Jy, the energy index of $\alpha$ = $0.97\pm0.04$, and the cut-off frequency of $\nu_c$ = $(1.1\pm0.4)\times 10^{13}$ Hz.
The evaluated FIR flux density of hot spot E ($F_{\nu, \mathrm{E-fit}}$) is about 10 per cent of the FIR flux density obtained from the aperture photometry (i.e. $F_{\nu,\mathrm{E-fit}}/F_\nu \sim 0.1$), except for at 70 \micron \ with $F_{\nu,\mathrm{E-fit}}/F_\nu$ = 0.3.
By subtracting the flux of hot spot E from the measured FIR flux ($F_\nu$), we estimated the FIR flux of hot spot D (i.e., $F_{\nu,\mathrm{D}}=F_\nu - F_{\nu,\mathrm{E-fit}}$).
We summarise $F_{\nu,\mathrm{D}}$ and $F_{\nu,\mathrm{E,fit}}$ in Table \ref{tab:flx}.
At the 160 \micron \ PACS band, the FIR fluxes of hot spots D and E are evaluated as $F_{\nu,\mathrm{D}}$ = 83$\pm$13 mJy and $F_{\nu,\mathrm{E-fit}}$ = 9.4$\pm$1.5 mJy.
As a result of the subtraction of the contamination from hot spot E, the derived FIR spectrum of hot spot D seems to smoothly connect to the radio spectrum within the errors, except for the 70 \micron \ band (see Figure \ref{fig:HS_D}).

\begin{table*}
  \begin{center}
    \caption{Estimated Flux densities of hot spot D and E.\label{tab:flx}}
    \begin{tabular}{lcccc}
      \hline
	\multirow{2}{*}{instrument} & $\lambda$ & $\nu$ & ${F_\mathrm{\nu, \mathrm{E-fit}}}^\mathrm{a}$ & ${F_{\nu,\mathrm{D}}}^\mathrm{b}$ \\
	& \micron & $10^{11}$ Hz & mJy & mJy \\
	    \hline
        SPIRE
      	& 350 & 8.6 & 22.0 $\pm$ 2.9 & 269 $\pm$ 66\\
      	& 250 & 12 & 15.4 $\pm$ 2.2 & 185 $\pm$ 58\\
      	PACS
      	& 160 & 19 & 9.4 $\pm$ 1.5 & 83  $\pm$ 13\\
      	& 100 & 40 & 5.3$\pm$ 1.0 & 45.4$\pm$ 9.4\\
		&  70 & 43  & 3.4$\pm$ 0.7 & 9.2 $\pm$ 7.5\\
      \hline
      \multicolumn{5}{l}{$a$ The FIR flux density of hot spot E calculated from the best-fit CPL model.}\\
      \multicolumn{5}{l}{$b$ The FIR flux density of hot spot D calculated as $F_\mathrm{D} = F_\mathrm{cor} - F_\mathrm{E,fit}$}\\
    \end{tabular}
  \end{center}
\end{table*}

\subsection{Spectral modelling of hot spot D}
\label{sec:spec_modeling}
\begin{figure}
    \includegraphics[width=\linewidth]{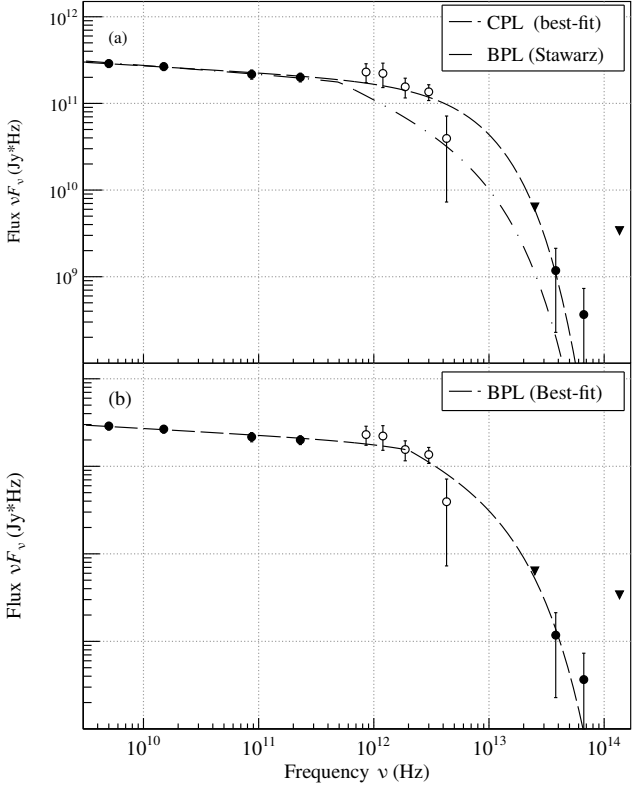}
\caption{SED of hot spot D. The open circles show the FIR spectrum of hot spot D ($F_\mathrm{D}$), after subtracting the contamination from hot spot E. The filled circles show the radio and NIR spectrum of hot spot D from literature. The dashed and dash-dotted line in panel (a) show the best-fit CPL model and the BPL model in \citealt{stawarz07}, respectively. The dashed line in panel (b) shows the best-fit BPL model.\label{fig:HS_D} \\ 
}
\end{figure}

We investigate the spectral properties of hot spot D.
Figure \ref{fig:HS_D} clearly shows that a simple PL does not reproduce the radio to NIR spectrum.
Therefore, we  first tried a CPL model to the observed spectrum, and derived the parameters as listed in Table \ref{tab:fit}.
Although  the CPL model, indicated with the dashed line in panel (a) of Figure \ref{fig:HS_D}, is statistically acceptable ($\chi^2$/d.o.f. = 6.78/8), we noticed several discrepancies between the observed and model spectra.
The best-fit CPL mode seems slightly flatter than the observed FIR spectrum of hot spot D, and hence, it possibly overestimates the 70 \micron \ flux.
This result suggests a spectral break in the FIR band.

In order to evaluate the possible break, we next adopted a broken PL model, subjected to a high energy cut-off written as 
\begin{equation}
F_{\nu,\mathrm{D}} \propto \left\{
        \begin{aligned}
            & \nu^{-\alpha}\exp(-\nu/\nu_\mathrm{c}) \ &\textrm{for} \ \nu < \ \nu_\mathrm{br}\\
            & \nu^{-(\alpha+\Delta \alpha)}\exp(-\nu/\nu_\mathrm{c}) \ &\textrm{for} \ \nu > \ \nu_\mathrm{br},
        \end{aligned}
        \right.
        \label{eq:bcpl}
\end{equation}
Hereafter, we simply refer to the model as the Broken PL (BPL) model.
We fitted the spectrum with the BPL model and found that the model improved the fitting.
However, due to the degeneracy between the change of the energy index ($\Delta \alpha$) and the break frequency ($\nu_\mathrm{br}$), it is difficult to determine the parameters simultaneously.
Thus, we searched for the range of $\Delta \alpha$ in which the fit is significantly improved.
As a result, we found that in the range of $0.23<\Delta \alpha < 7$, the model reduces the chi-square by at least 1 in comparison to the CPL model.
Therefore, we fixed the index change at $\Delta \alpha$ = 0.5 to constrain the break frequency, since it is theoretically consistent with the index change predicted from the diffusive shock acceleration under the continuous energy injection, accompanied with a radiative cooling \citep{heavens87,carilli91}.
\cite{stawarz07} interpreted the spectrum of hot spot D, without the FIR data, by the BPL model with $\Delta \alpha$ = 0.5, and derived the parameters denoted as "BPL (Stawarz)" in Table \ref{tab:fit}.
Although the BPL model with the parameters by \cite{stawarz07} agrees with the radio data, it significantly underestimates the FIR spectrum, as shown with the dash-dotted line in panel (a) of Figure \ref{fig:HS_D}.
This suggests that the break frequency is higher than their result ($\nu_\mathrm{br} = 0.5 \times 10^{12}$ Hz; \citealt{stawarz07}).

To determine precisely the break frequency, we re-fitted the BPL model  to the spectrum of hot spot D, by including the FIR data.
The dashed line in panel (b) in Figure \ref{fig:HS_D} shows the best-fit BPL model ($\chi^{2}$/d.o.f. = 4.8/7), which successfully reproduces the overall spectrum in the radio, FIR and NIR ranges.
The best-fit parameters are summarised in Table \ref{tab:fit}.
By adopting the BPL model, the fit is slightly improved compared to the CPL model; the null-hypothesis probability of the \textit{F}-test between the best-fit CPL and BPL is 14 per cent.
The best-fit break frequency is determined as $\nu_\mathrm{br} = 2.0^{+1.2}_{-0.8}\times10^{12}$ Hz.
The derived break frequency is significantly higher than that adopted in \cite{stawarz07}.
The other parameters are found to stay almost unchanged from those of the CPL model and those of the BPL model in \cite{stawarz07}.
Thanks to the FIR data, we have succeeded in determining the break frequency of hot spot D for the first time.

\begin{table*}
  \begin{center}
    \caption{Best-fit parameters to the synchrotron spectrum  of hot spot D.\label{tab:fit}}
    \begin{tabular}{lccccc}
      \hline
         \multirow{2}{*}{model} & ${F_\nu (\mathrm{10 \ GHz})}^\mathrm{a}$ & \multirow{2}{*}{${\alpha}^\mathrm{b}$} & ${\nu_\mathrm{br}}^\mathrm{c}$ & ${\nu_\mathrm{c}}^\mathrm{d}$ & \multirow{2}{*}{$\chi^2$/d.o.f.}\\
         & Jy & & Hz & Hz \\
        \hline
        \vspace{-8pt}\\
        CPL & 27$\pm$1.7 & 1.06$\pm$0.03& --- & (0.8$\pm$0.1)$\times$10$^{13}$ & 6.7/8\\
        BPL (Stawarz)$^\mathrm{e}$ & 27 & 1.1 & 0.5$\times$10$^{12}$ & 0.9$\times$10$^{13}$ & 20/11 \\
        BPL$^\mathrm{e}$ & 27$\pm$1.8 & 1.07$\pm$0.03 & 2.0$^{+1.2}_{-0.8}\times$10$^{12}$ & (1.2$\pm$0.3)$\times$10$^{13}$ & 4.8/7 
        \vspace{2pt}\\
      \hline
      \multicolumn{6}{l}{a The flux density at 10 GHz}\\
      \multicolumn{6}{l}{b The energy index at lower frequency than the break and cut-off}\\
      \multicolumn{6}{l}{c The break frequency of the BPL model}\\
      \multicolumn{6}{l}{d The cut-off frequency}\\
      \multicolumn{6}{l}{e The energy-index change is fixed as $\mathit{\Delta\alpha}=0.5$.}\\
    \end{tabular}
  \end{center}
\end{table*}

\section{Discussion} 
\subsection{FIR Evidence of the cooling break}  
\label{sec:FIR_break}
We detected the FIR emission associated with hot spot D, located where the eastern jet of the radio galaxy Cygnus~A terminates, in the 70-350 \micron \ images obtained by the \textit{Herschel}/SPIRE and PACS.
This is the second hot spot detected in the FIR range after the west hot spot of Pictor A \citep{isobe20}.
After subtracting the contamination from hot spot E, the FIR spectrum of hot spot D is revealed to be consistent with the extrapolation from the radio PL spectrum.
Therefore, the FIR emission is naturally attributed to the synchrotron emission from the same electron population producing for the radio emission.

In the radio to FIR range, the spectral energy index is derived as $\alpha \simeq 1.1$.
Although the index is explainable by the diffusive shock acceleration \citep{bell78}, it is larger than the canonical value, $\alpha=0.5$, corresponding to the strong non-relativistic shock. 
Several possible ideas for the spectral softening are presented in \cite{stawarz07},
including some relativistic effects and/or magnetic-field configurations,
and thus, we did not discuss this issue any further in the present paper.
The synchrotron spectral shape at the higher frequency end depends on physical conditions associated with the acceleration region. 
Several models, including a simple one-shot energy-injection one, exhibit a high-frequency cut-off at the frequency corresponding to the maximum energy of the accelerated electrons.
In contrast, the diffusive shock acceleration under a continuous injection \citep{meisenheimer89,carilli91} predicts a spectral break with $\Delta \alpha = 0.5$ due to radiative cooling, in addition to the high-frequency cut-off.
By filling the gap between the radio and NIR bands with the FIR data, we, for the first time, confirmed that the BPL model with $\Delta \alpha = 0.5$ better reproduces the observed synchrotron spectrum of the object than the CPL one.
This indicates that diffusive shock acceleration under continuous energy injection is the dominant acceleration process in hot spot D.

Previous studies tried to estimate the cooling break frequency of hot spot D, though they were not conclusive.
\cite{carilli91} proposed the break feature in the GHz range as the cooling break.
However, theoretical studies found that a cooling break at higher frequency than GHz is preferred in hot spots \citep[e.g.,][]{kino04}.
In addition, \cite{stawarz07} clearly demonstrated that the GHz break feature does not represent the cooling break, since they precisely evaluated the spectral index change at the GHz feature as $\Delta \alpha \simeq 0.8$ , which is inconsistent with the cooling break (i.e., $\Delta \alpha = 0.5$), by re-analysing the high-resolution radio data.
Instead, \cite{stawarz07} assumed a cooling break at the frequency of $\nu_{\rm br} = 0.5 \times 10^{12}$ Hz to reproduce the spectrum they obtained.
However, they were unable to justify this assumption because the frequency falls in the spectral gap between the radio and NIR ranges. 
We overcame this difficulty by introducing the newly obtained FIR data with \textit{Herschel}.
Thus, we have succeeded in directly measuring the cooling break as $\nu_{\rm br} = 2.0^{+1.2}_{-0.8}\times 10^{12}$ Hz.

\subsection{Magnetic-field estimation} 
\label{sec:B_estimate}
In this section, we evaluate the magnetic field strength $B$ in hot spot D.
First, we constrain the magnetic field strength $B$ as a function of the observed radius of the hot spot, $R$.
Next, we re-evaluate the observed X-ray spectrum with the Synchrotron-Self-Compton (SSC) process 
to obtain an independent constraint on $B$ and $R$ from the cooling break.
Finally, by combining these two considerations, 
we precisely determine the magnetic field strength in hot spot D. 

\subsubsection{Constraint from the cooling break frequency}
\label{sec:B_from_break}
From the cooling break frequency, we measure the magnetic field strength $B$ as the function of the radius $R$.
The cooling break is determined 
by mutual balance between an electron radiative cooling time scale, 
$t_\mathrm{syn} = \frac{6\pi m_\mathrm{e} c}{B^2\sigma_\mathrm{T}\gamma}$ 
and an adiabatic loss time scale, $t_\mathrm{ad} = \frac{2R\eta}{\beta c}$,
where  $m_\mathrm{e}$, $c$, $\sigma_\mathrm{T}$, $\gamma$ and $\beta$ are 
the electron rest mass, speed of light, Thomson cross section, 
electron Lorentz factor, and downstream flow velocity 
behind the shock in the shock frame.
The parameter $\eta$ is a correction factor from the physical to effective lengths 
along the jet (i.e., $0 < \eta \le 1$).
This parameter simply quantifies non-uniformity in the post-shock region,
and is equivalent to the filling factor of the magnetic field. 
Here, for simplicity, the Compton cooling is neglected based on the relative observed strength of the synchrotron and SSC components.
Because the two time scale becomes equal to each other at the cooling break,
the Lorentz factor at the cooling break is given by
\begin{equation}
    \gamma_\mathrm{br}=\frac{3\pi m_\mathrm{e} c^2 \beta}{\sigma_\mathrm{T} \eta RB^2},
    \label{eq:break}
\end{equation}
\citep{inoue96}.
For high energy electrons above the cooling break, $\gamma > \gamma_\mathrm{br}$,
the synchrotron spectral slope changes by $\Delta \alpha = 0.5$,
assuming that the energy injection is constant in time.
The synchrotron frequency corresponding to the break Lorentz factor is derived 
as $\nu_{\mathrm{br}}= \frac{3eB\gamma_\mathrm{br}^2}{4\pi m_\mathrm{e}c}$,
where $e$ is the elementary charge.
Thus, the magnetic field $B$ is estimated from the break frequency as
\begin{eqnarray}
B &=& \left( \frac{27\pi m_\mathrm{e} e c^3 \beta^2}{4\sigma_\mathrm{T}^2} \eta^{-2} R^{-2} 
    \nu_\mathrm{br}^{-1} \right)^{1/3} \notag \\
 &\simeq& 190~\mathrm{\mu G}
    \times \left(\frac{\beta}{1/3}\right)^{\frac{2}{3}} 
    \eta^{-\frac{2}{3}} 
    \left(\frac{R}{1 \ \mathrm{kpc}}\right)^{-\frac{2}{3}} 
    \left(\frac{\nu_\mathrm{br}}{10^{12} \mathrm{Hz}}\right)^{-\frac{1}{3}},
\label{eq:B-R}
\end{eqnarray}
This method was successfully applied to estimate the magnetic field 
in blazars \citep[e.g.,][]{inoue96,kataoka00} and hot spots, 
including the west hot spot of Pictor A \citep{isobe20}.

Here, as a baseline scenario,
we adopted a uniform one-zone model ($\eta = 1$) 
with an ideal strong relativistic shock \citep[$\beta=1/3$; e.g., ][]{kirk99}.
By substituting the observed break frequency,
$\nu_\mathrm{br}=2.0^{+1.2}_{-0.8}\times10^{12}$ Hz, in Equation \ref{eq:B-R}, 
we derive the magnetic field as a function of the radius as,
\begin{equation}
B \simeq 150_{-20}^{+30} \times \left(\frac{R}{1 \ \mathrm{kpc}}\right)^{-\frac{2}{3}} \mathrm{\mu G}.
\label{eq:B}
\end{equation}
The area enclosed by the thick blue lines in panel (a) of Figure \ref{fig:BvsR} indicates 
the acceptable magnetic field given by Equation \ref{eq:B}.

In order to further constrain the magnetic field from Equation \ref{eq:B}, 
we evaluate the radius of the object from the previous studies \citep[e.g.,][]{harris94,wilson00,kino04,stawarz07}.
We conservatively adopt the radius range of $0.8$ kpc $<R< 1.6$ kpc,
which roughly covers the radius estimations in the previous studies. 
The lower limit corresponds to the radius adopted in \cite{stawarz07} 
based on the VLA high-resolution observation \citep{perley84},
while the upper one was derived from the X-ray image with \textit{Chandra}
\citep{wilson00}.
The horizontally hatched area in panel (a) of Figure \ref{fig:BvsR} shows the magnetic field for the adopted radius range. 
Thus, the magnetic field is determined as $B=90$--$210$ $\mathrm{\mu G}$ 
by using the cooling break.

\subsubsection{Constraint from the X-ray spectrum}
\label{sec:B_from_SSC}
In order to constrain the magnetic field independently from the cooling break.
we made a Synchrotron-Self-Compton (SSC) model \citep{band85} for the observed X-ray spectrum
of hot spot D.
The SSC model is widely adopted to interpret the X-ray spectra of numbers of hot spots,
and to constrain their magnetic field 
\citep{wilson00,hardcastle04,kino04,kataoka05}.
In the SSC process, the flux ratio of the synchrotron to SSC components 
depends on the magnetic field and the radius of the source.
\cite{stawarz07} estimated the magnetic field of hot spot D 
as $B=270$ $\mathrm{\mu G}$,
by applying the SSC model to its X-ray spectrum for the fixed radius of $R$=0.8 kpc.
We, here, re-modelled the X-ray spectrum of the object 
by the SSC process to constrain the magnetic field 
for the radius range of 0.8 kpc <$R$< 1.6 kpc, as we adopted in the previous subsection.

\begin{figure*}
  \begin{center}
    \includegraphics[width=\linewidth]{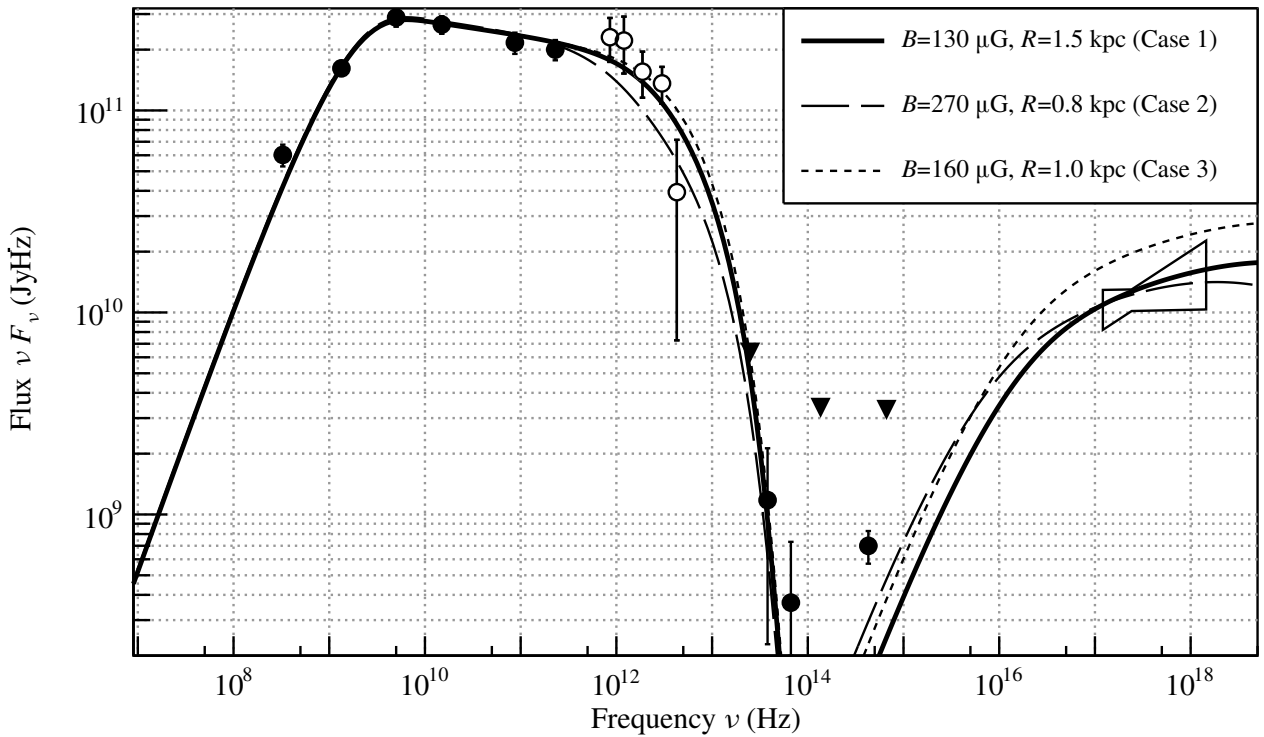}
    \caption{The broadband SED of hot spot D and the SSC model calculated in section 4.2.2.
    The open circles indicate the FIR spectrum obtained in this work.
    The black filled circles and the black filled triangles are the data points and upper limits (\citealt{stawarz07} and references there in).
    The solid, dashed and dotted lines indicate the synchrotron+SSC models for Cases 1, 2, and 3, respectively.
    \label{fig:SSC}}
  \end{center}
\end{figure*}

We compiled the broadband SED
of hot spot D, as shown in Figure \ref{fig:SSC}.
The FIR data are those we obtained in section \ref{sec:obs_ana} (See Table \ref{tab:flx}).
The radio, NIR, and X-ray data are taken from \cite{stawarz07},
in which hot spot D is resolved from hot spot E 
with the high-resolution images.
The synchrotron spectrum of this object is measured 
in a wide frequency range between the radio and NIR bands.
As mentioned in section \ref{sec:FIR_break}, the spectrum appears to exhibit a low-frequency cut-off feature around 10$^9$ Hz.
Although the origin of this feature is under debate 
\citep{carilli91,mckean16,stawarz07},
we do not discuss it any further in the present paper.
Thus, we only examine the synchrotron spectrum 
above the minimum frequency of $\nu_\mathrm{min} = 1.4 \times 10^{9}$ Hz. 
In contrast, the synchrotron spectrum shows the high frequency cut-off,
of which the frequency is determined as $\nu_\mathrm{cut} = (1.2\pm0.3)\times10^{13}$ Hz, 
in addition to the cooling break, as discussed in section \ref{sec:spec_modeling}.
This cut-off is attributed to the maximum electron energy.
The X-ray spectrum significantly exceeds the extrapolation of the radio-to-NIR synchrotron spectrum.
Therefore it is thought to be produced via the SSC process. 

We calculated the synchrotron and SSC spectra to reproduce the observed radio-to-X-ray SED.
For the calculation, the open source package of {\small NAIMA versioin 0.9.1} \citep{zabalza15} was utilised.
Based on the observed shape of the synchrotron spectrum in the range of $\nu > \nu_\textrm{min}$,
the input electron spectrum is assumed to be simply described 
by a broken power-law model as shown below;
\begin{equation}
N_\mathrm{e}({\gamma}) = N_0\times \left\{
        \begin{aligned}
            & \left(\frac{\gamma}{\gamma_\mathrm{min}} \right)^{-p}& \ &\textrm{for} \ \gamma_\mathrm{min} \leq \gamma < \ \gamma_\mathrm{br}\\
            & \left(\frac{\gamma_\mathrm{br}}{\gamma_\mathrm{min}} \right)^{-p} \times \left(\frac{\gamma}{\gamma_\mathrm{br}}\right)^{-(p+1)}&  &\textrm{for} \ \gamma_\mathrm{br}  \leq \gamma \leq \gamma_\mathrm{max}\\
            &\ \ \ \ \ 0& &\mathrm{ otherwise,}
        \end{aligned} 
        \right.
        \label{eq:electron_dist}
\end{equation}
where $N_0$, $p$, $\gamma_\mathrm{min}$, and $\gamma_\mathrm{max}$ 
are the normalisation in the unit of the electron number per unit Lorentz factor,
the spectral energy index, the minimum and maximum electron Lorentz factors, respectively.
Thus, it is necessary to specify the seven parameters to simulated the overall spectrum; i.e., 
$B$, $R$, $N_0$, $p$, $\gamma_\mathrm{min}$, $\gamma_\mathrm{br}$ and $\gamma_\mathrm{max}$.

Here, 
the SSC model calculation is performed for our baseline scenario, 
while the impact of the parameter $\eta$ is discussed in \S \ref{sec:uncertainty}.
Through the following procedures, 
we check if a certain pair of $B$ and $R$ is able to reproduce the observed SED shown in Figure \ref{fig:SSC}.
Based on the synchrotron energy index of $\alpha=1.07$ between the radio and FIR bands
determined in section \ref{sec:spec_modeling},
the electron spectral index of $p=2\alpha+1=3.14$ is adopted.
The minimum and cut-off synchrotron frequencies, 
$\nu_\mathrm{min}$ and $\nu_\mathrm{cut}$ respectively,
are transformed into the minimum and maximum Lorentz factors, 
$\gamma_\mathrm{min}$ and $\gamma_\mathrm{max}$, 
as $\gamma=\sqrt{\frac{4\pi m_\mathrm{e}c \nu }{3qB}}$.
In order to keep consistency to $B$ and $R$,
we adopted the break Lorentz factor calculated from Equation \ref{eq:break}
instead of the observed value.
We determine the electron normalisation $N_0$
so that the simulated synchrotron flux reproduces 
the observed one as $F_\mathrm{10~GHz}=27$ Jy at 10 GHz \citep{stawarz07}.
With these parameters, we compared the SSC model spectrum with the observed X-ray spectrum.
If the calculated flux agrees with the observed X-ray one at 1 keV,
$F_\mathrm{X}=47.0\pm5.9$ nJy \citep{stawarz07},
the input $B$-$R$ pair is regarded as viable. 
We iterate these procedures for a wide range on the $B$-$R$ plane.
We show the synchrotron and SSC spectra for some representative values of $B$ and $R$ (i.e., Cases 1--3) in Figure \ref{fig:SSC} and tabulated the parameters in Table \ref{tab:SSC_par}.

\begin{figure*}
  \begin{center}
        \includegraphics[width=\linewidth]{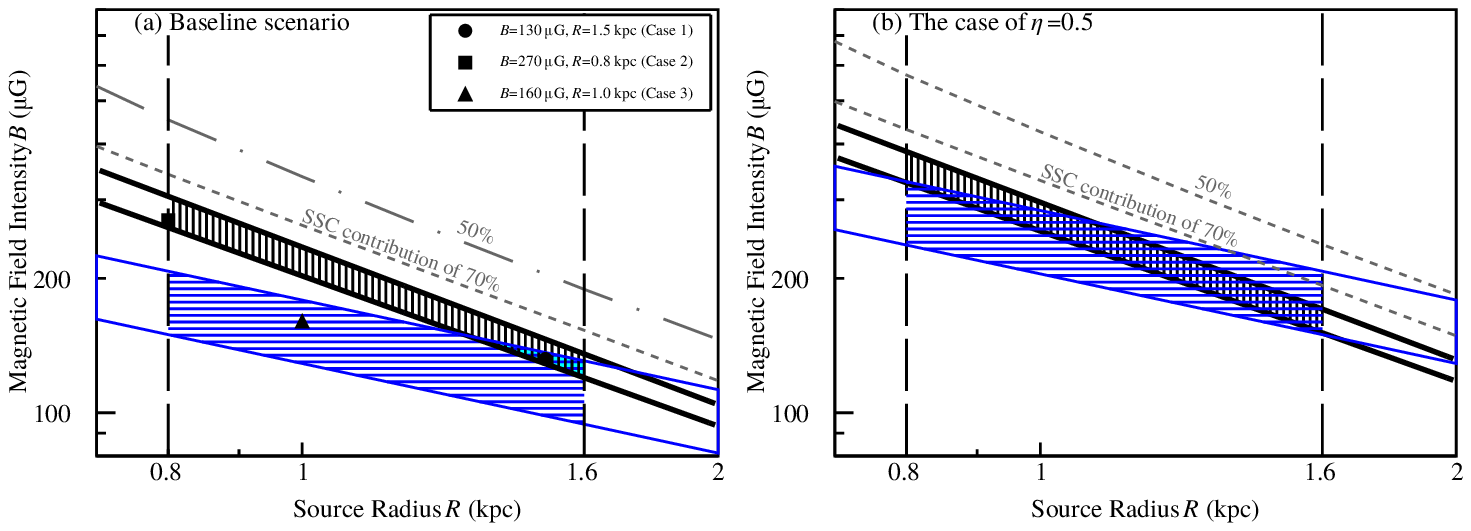}
    \caption{The magnetic field $B$ and the radius $R$ for hot spot D.
    The vertical dashed lines indicate the radius of $R=0.8$ and $1.6$ kpc.
    The constraints for our baseline scenario 
    ($\eta = 1 $ and $\beta = 1/3$) are displayed in panel (a), 
    while those for $\eta = 0.5$ are plotted in panel (b) 
    to visualise the sensitivity of the result on $\eta$ (see section \ref{sec:uncertainty}).
    The region enclosed by the blue solid line is the constraint from the FIR cooling break.
    Taking into account the previous results on the radius 
    \citep[the dashed lines;][]{perley84,wilson00}, the acceptable region is limited to the hatched region with the blue lines.
    The region hatched by the black vertical lines  show the area matched to the SSC constraint with the radius condition considered.
    The cyan filled region indicates the parameters satisfying simultaneously the constraint from the cooling break and that from the SSC X-ray flux.
    The filled circle, filled square, and the filled triangle indicate Cases 1, 2, and 3, respectively, of which the parameters are listed in Table \ref{tab:SSC_par}.
    The grey dash-dotted and dotted lines indicate the conditions where the SSC flux reproduces 50 and 70 per cent of observed X-ray flux, respectively.
    \label{fig:BvsR}}
  \end{center}
\end{figure*}

The area enclosed by the two thick black solid lines in Figure \ref{fig:BvsR} 
indicates the acceptable values of the magnetic field and radius.
As discussed in Section \ref{sec:B_from_break},
we adopted the additional constraint on the radius as $0.8$ kpc $<R< 1.6$ kpc.
Thus, we finally obtained the black vertically hatched area in panel (a) of Figure \ref{fig:BvsR},
within which the observed 10 GHz radio and 1 keV X-ray fluxes 
are consistently described with the synchrotron and SSC components, respectively.

\subsubsection{Magnetic field determination}
By combining the investigation into the cooling break and 
the SSC modelling of the X-ray spectrum, 
shown in sections \ref{sec:B_from_break} and \ref{sec:B_from_SSC} respectively,
we tightly constrained the magnetic field strength in hot spot D, 
for our baseline scenario.
The derived $B$-$R$ conditions from the individual considerations 
overlap with each other in the blue filled region in Figure \ref{fig:BvsR}, covering magnetic field strength of $B=120$--$150$ $\mathrm{\mu G}$ 
for radii in the range $R=1.3$--$1.6$ kpc.
This is the most stringent magnetic-field constraint ever achieved for hot spots of radio galaxies
\citep[e.g.,][]{hardcastle04,kataoka05}.

In order to visually validate the magnetic-field estimation, 
the synchrotron and SSC model spectra are plotted in Figure \ref{fig:SSC} for three representative cases (Cases 1, 2, and 3).
As shown with the filled circle in panel (a) of Figure \ref{fig:BvsR}, 
Case 1 ($B$=130 $\mathrm{\mu G}$ and $R$=1.5 kpc) simultaneously 
satisfies the two constraints. 
Therefore, the observed cooling break and X-ray flux 
are successfully reproduced by the synchrotron and SSC model components, respectively, 
as displayed with the thick solid lines in Figure \ref{fig:SSC}.
Case 2 ($B$=270 $\mathrm{\mu G}$ and $R$=0.8 kpc), 
plotted with the filled box in panel (a) of Figure \ref{fig:BvsR}, 
is located within the acceptable region for the X-ray spectrum,
but outside the region for the cooling break. 
The parameters of Case 2 were adopted in \cite{stawarz07}. 
The dashed line in Figure \ref{fig:SSC} clearly shows that 
the SSC model flux agrees with the observed X-ray one, 
although the cooling break frequency predicted by the model 
is lower than the observed value, causing the model synchrotron spectrum to fall below the observed spectrum in the FIR.
The filled triangle in panel (a) of Figure \ref{fig:BvsR} points to
the parameters of Case 3 ($B=160$ $\mathrm{\mu G}$ and $R=1.0$ kpc),
which only meets the cooling break condition.
Thus, the simulated synchrotron spectrum is compatible with the observed cooling break,
though the SSC model overestimates the X-ray flux,
as depicted with the dashed line in Figure \ref{fig:SSC}. 

With the cooling break properly taken into account in the SSC model,
we have succeeded in strongly constraining the magnetic field in hot spot D of Cygnus A ($B$=120--150 $\mathrm{\mu G}$).
If this method is systematically applied to the other hot spots, 
their magnetic field is expected to be precisely measured. 
In fact, a number of hot spots are suggested to exhibit the cooling break in the FIR range,
because their magnetic field and radius are typically evaluated 
in the range of $B= 100$--$500$ $\mathrm{\mu G}$ and $R=0.3$--$3$ kpc 
\citep[e.g][]{kataoka05,zhang18}.
In addition, \cite{cheung05} predicted that 
the cooling break is located around the FIR band for three optical hot spots, 
by simply connecting the radio and optical spectra with the broken PL model.
Therefore, future FIR studies gives a powerful tool to evaluate the physical condition in the hot spots, 
by detecting the cooling break.

\begin{table*}
  \begin{center}
    \caption{Parameters for the calculation of the synchrotron and SSC spectra \label{tab:SSC_par}}
    \begin{tabular}{lccccccc}
      \hline
          \multirow{2}{*}{case} & $B$ & $R$ & \multirow{2}{*}{$p$} &$N_\mathrm{0}$ &\multirow{2}{*}{$\gamma_\mathrm{min}$} & \multirow{2}{*}{$\gamma_\mathrm{max}$} & \multirow{2}{*}{$\gamma_\mathrm{br}$}\\
         & $\mathrm{\mu G}$ & kpc & & particle per Lorentz factor &  & & \\
        \hline
        \vspace{-8pt}\\
        Case 1 & 130 & 1.5 & 3.14 & $6.4\times10^{56}$ & 1.5$\times$10$^{3}$ & 1.4$\times$10$^{5}$ & 4.9$\times$10$^{4}$\\
        Case 2 & 270 & 0.8 & 3.14 & $4.4\times10^{56}$ & 1.1$\times$10$^{3}$ & 9.9$\times$10$^{4}$ & 2.1$\times$10$^{4}$\\
        Case 3 & 160 & 1.0 & 3.14 & $5.7\times10^{56}$ & 1.4$\times$10$^{3}$ & 1.3$\times$10$^{5}$ & 4.9$\times$10$^{4}$\\
      \hline
    \end{tabular}
  \end{center}
\end{table*}

\subsubsection{Energetics}
\label{sec:energetics}

We evaluated the energy densities of the non-thermal electrons and magnetic field, because these two components are directly accessible through the synchrotron radio an SSC X-ray radiations.
Here, the physical parameters for Case 1 (see Table \ref{tab:SSC_par}) are adopted,
since the case satisfied both the cooling break and SSC constraints.
The energy density of non-thermal electrons is calculated by integrating Equation \ref{eq:electron_dist} from $\gamma_\mathrm{min}$ to $\gamma_\mathrm{max}$ as 
$U_\mathrm{NT,e}=m_\mathrm{e}c^2\int_{\gamma_\mathrm{min}}^{\gamma_\mathrm{max}} \gamma N_\mathrm{e}(\gamma) \mathrm{d}\gamma \approx 2.9\times10^{-9}$ erg cm$^{-3}$.
The energy density of the magnetic field is evaluated as $U_B=B^{2}/8\pi = 0.67\times10^{-9}$ erg cm$^{-3}$.
The derived $U_\mathrm{NT, e}$ and $U_{B}$ values at hot spot D are consistent with the previous estimations by \cite{stawarz07, snios18}.
Thus, an non-thermal-electron dominance of $U_\mathrm{NT,e}/U_B\sim4$ is indicated in hot spot D.
A similar non-thermal-electron dominance is reported for the hot spot A, which is located at the west-jet terminal of Cygnus~A \citep{kino04}.

\subsubsection{Possible uncertainties}
\label{sec:uncertainty}
In this subsection, 
we evaluate the influence of possible deviations from our baseline scenario,
adopted in sections \ref{sec:B_from_break} -- \ref{sec:energetics}.
As is clear from equation (\ref{eq:B-R}), 
the magnetic field estimated from the cooling break is dependent on 
the adopted values of $\eta$ and $\beta$.
In addition, the parameter $\eta$ has a possible impact on the SSC calculation, 
since it is related to the effective volume of the emission region.
From the observational point of view, 
it is difficult to specify these two parameters.
Therefore, we investigate the sensitivity of the magnetic field estimation 
to $\eta$ and $\beta$.

Firstly, we deal with the effect of $\eta$, 
which is adopted to parameterize the post-shock non-uniformity.
Numerical studies \citep[e.g.,][]{inoue09} frequently imply 
a post-shock inhomogeneous condition with a highly entangled magnetic field.
Such a non-uniformity, described as $\eta < 1$, 
is expected to enhance the magnetic field estimation 
via the cooling break, 
since equation (\ref{eq:B-R}) indicates $B\propto \eta^{-2/3}$.
As an example of highly non-uniform cases, 
the magnetic field for $\eta = 0.5$
is plotted on panel (b) of Figure \ref{fig:BvsR} 
with the blue hatched region.
In comparison to our baseline scenario ($\eta=1$)
shown in panel (a) of Figure \ref{fig:BvsR}, 
the compatible region to the observed cooling break is shifted upward 
by a factor of $\sim1.6$ on the $B$--$R$ plane.
In contrast, the effective volume of the emission region is proportional to $\eta$,
on condition that only the non-uniformity along the jet 
is taken into consideration for simplicity. 
As a result, the magnetic field to reproduce simultaneously 
the observed radio and X-ray fluxes is thought to scale as $B\propto \eta^{-1/3}$.
As shown with the black vertically hatched area in panel (b) of Figure \ref{fig:BvsR},
the magnetic field for $\eta = 0.5$ obtained from the SSC modelling is predicted to be enlarged 
by a factor of $\sim 1.3$, compared with that for our baseline scenario.
By combining these two considerations,
it is found that a slightly wider range of the magnetic field 
as $B=150$--$310$ \ $\mathrm{\mu G}$ becomes acceptable.

Secondly, we make a brief comment on the impact of the $\beta$, 
i.e., the down-flow velocity in the shock frame normalised to the light speed.
The impact of $\beta$ on the magnetic field estimation 
seems to be more straight forward than that of $\eta$,
because $\beta$ only affects the cooling break condition 
as $B\propto \beta^{2/3}$ (see equation (\ref{eq:B-R})).
By adopting $\beta > 1/3$ (e.g., in the case of the weak shocks), 
a higher magnetic-field value becomes acceptable, 
since only the blue hatched region moves upward 
in panel (a) of Figure \ref{fig:BvsR}.
However, the $\beta$ value significantly smaller 
than $1/3$, corresponding to such as non-relativistic cases, is expected to be unrealistic, 
since it extinguishes the overlap between the cooling-break and SSC constraints 
on the $B$-$R$ diagram, 
unless a significant non-uniformity is simultaneously considered.

\subsection{Validity of the SSC scenario}
When we estimated the magnetic field strength of hot spot D 
in Section \ref{sec:B_estimate},
we simply assumed that all the observed X-ray flux is attributable 
to the SSC emission.
Strictly speaking, there is no physical rationale for this assumption,
although it is widely applied to X-ray studies of hot spots 
\citep{hardcastle04,kataoka05,zhang18}.
Actually, it is suggested that 
the X-ray spectrum of some optical hot spots is contaminated 
by other spectral components, including the synchrotron emission 
from an additional electron population
\citep{wilson01,hardcastle07,kraft07,perlman10}.
As a by-product of the cooling break determined from the FIR data, 
we successfully quantitatively restrict 
the contribution of components different from the SSC one
to the observed X-ray spectrum of hot spot D.

Compact synchrotron emitting sources, including the hot spots, 
are inevitably accompanied by some SSC emission.
However, to determine the SSC flux, the observed X-ray spectrum may need to be corrected for contamination by other possible X-ray emission processes in the hot spot. 
The SSC model flux is predicted to scale, roughly, as $F_\mathrm{SSC} \propto B^{-2}$, when the synchrotron
flux is held fixed. 
Therefore, increasing the fraction of the contaminating spectral component requires the magnetic field to be enhanced and, hence, the $B-R$ region in Figure \ref{fig:BvsR} allowed by the SSC model (i.e., the vertically hatched area) is shifted upward.

The dotted and dash-dotted lines in Figure \ref{fig:BvsR} represent the $B$-$R$ relations, 
derived in a manner similar to that adopted in Section \ref{sec:B_from_SSC},
for the cases where the SSC model reproduces 
$70$ and $50$ per cent of the observed X-ray flux, respectively.
Such a small SSC contribution is clearly rejected in the case of our baseline scenario
from panel (a) of Figure \ref{fig:BvsR}, 
because  these two lines do not intersect the horizontally hatched region 
obtained by the cooling break consideration.
In order for the regions from the cooling break and SSC conditions to overlap with each other,
we found that the SSC fraction of higher than
$\sim98$ per cent of the best-fit X-ray flux (i.e., 47 nJy) is required. 
In the case of the lowest acceptable SSC fraction of 98 per cent,
the regions from the two conditions cross at the top-right corner 
of the horizontally hatched region in panel (a) of Figure \ref{fig:BvsR}, 
which corresponds to the magnetic field of $B=130$ $\mathrm{\mu G}$ 
and the radius of $R=1.6$ kpc, respectively.
Even if the non-uniformity discussed in section \ref{sec:uncertainty} is assumed,
the SSC dominance in the observed X-ray flux is still justified for a wide range of $\eta$, 
e.g., the SSC contribution of $\gtrsim 60$ per cent for $\eta=0.5$ 
as shown in panel (b) of Figure \ref{fig:BvsR}.

In this way, we have put a very tight lower limit on the SSC contribution 
to the X-ray spectrum of hot spot D in the radio galaxy Cygnus A, 
owing to the detection of the cooling break.
This result strongly indicates that 
the SSC emission significantly dominates the X-ray spectrum of this object.
If the cooling break is systematically detected in the other hot spots, 
the SSC scenario for the hot spots' X-ray spectrum is possibly justified with high reliability. 

\section{SUMMARY}
By making use of the \textit{Herschel} SPIRE and PACS data, 
we have for the first time detected the FIR source associated with 
hot spot D in the radio galaxy Cygnus A.
The important results and discussion derived 
from the FIR data are summarised below.

\begin{enumerate}[1.]
\item The spatial analysis revealed that the FIR source is 
extended with a PSF-deconvolved size of $d_{\rm src} = 8 \pm 2$ arcsec,
and its peak position is shifted by $\sim 4$ arcsec from hot spot D towards hot spot E.
These properties indicate a significant contamination from hot spot E.
Thus, we first performed SPIRE and PACS photometry of the FIR source 
with a circle covering both hot spots D and E. 
The FIR of the source is evaluated as 
$291 \pm 66$ mJy and $92.3 \pm 13$ mJy 
at 350 \micron\ and 160 \micron, respectively.

\item Since the FIR spectrum of the source is found 
to slightly exceed the extrapolation of the radio spectrum 
of hot spot D, the contamination from hot spot E 
is spectroscopically confirmed. 
By interpolating the radio and NIR spectra,
the FIR flux of hot spot E was evaluated 
as $\sim10$ per cent of the total FIR source flux.
By subtracting the estimated FIR flux of hot spot E, 
the FIR flux of hot spot D was evaluated 
as $269\pm66$ mJy and $83\pm13$ mJy at 350 \micron\ and 160 \micron,
smoothly connecting the radio and NIR spectra of hot spot D.
\item  The radio-to-NIR synchrotron spectrum of hot spot D is indicated 
to prefer the BPL model to the CPL one.
The change in the spectral index of the BPL model
becomes consistent with the prediction from 
the diffusive shock acceleration under the
continuous energy injection, accompanied with a radiative cooling,
i.e., $\Delta \alpha=0.5$.
Thanks to the FIR data,
the break frequency was derived as $\nu_\mathrm{br}=2.0^{+1.2}_{-0.8}\times 10^{12}$ Hz, with the index change fixed at $\Delta \alpha=0.5$. 
The break is naturally interpreted as the cooling break.
\item 
Based on the cooling break interpretation,
the derived break frequency of hot spot D 
is converted into the magnetic field strength for
the uniform one-zone model with the strong relativistic shock 
(i.e. $\eta=1$ and $\beta=1/3$). 
In addition to this estimation, 
we also evaluated the magnetic field strength 
by assuming that  all the observed X-ray spectrum 
($F_\mathrm{X}=47.0 \pm 5.9$ nJy at 1keV; \cite[][]{stawarz07}) 
is attributable to the SSC process.
By combining the two constraints, 
we determined the magnetic field intensity of hot spot D 
as $B=120$--$150 \ \mathrm{\mu G}$ 
for the radius of $R=1.3$--$1.6$ kpc (panel a of Figure \ref{fig:BvsR}). 
This magnetic field strength indicates 
a non-thermal electron dominance 
with the electron-to-magnetic-field energy density ratio 
of $U_\mathrm{NT,e}/U_\mathrm{B}\sim 4$. 
\item 
We discussed the impact of the possible deviation 
from the uniform one-zone model with the strong shock.
If we adopted $\eta = 0.5$ for a representative case 
of a significant non-uniformity, 
the acceptable magnetic field strength is expected to 
become slightly higher as $B=150$--$310 \ \mathrm{\mu G}$,
since the magnetic field from the cooling break scales 
as $\propto \eta^{-2/3}$ and that 
from the SSC model as $\propto \eta^{-1/3}$. 
The $\beta$ value significantly smaller than $1/3$ is regarded 
as unrealistic, 
because the region which simultaneously meets 
the cooling-break and SSC constraints is expected to disappear.
\item 
If the observed X-ray flux is contaminated by emission processes
different from the SSC one, 
the acceptable magnetic field is expected to be higher
since the SSC flux is thought to scale as $F_\mathrm{SSC} \propto B^{-2}$ 
for the fixed synchrotron flux.
However, a detailed investigation into 
the cooling-break and SSC conditions indicates 
that the SSC fraction to the observed X-ray flux 
is higher than $\sim 98$ per cent in the case of 
the uniform one-zone model accompanied by the strong shock.
Even if the deviation from this ideal assumption is considered,
a high SSC dominance is found to be justified for a wide range of $\eta$ 
(e.g., the SSC fraction of $>60$ per cent for $\eta=0.5$--$1$). 
\end{enumerate}
We have successfully constrained the physical parameters of hot spot D, especially magnetic field intensity, by determining the cooling break frequency in the FIR band. 
Based on this result, 
FIR observations are expected to be helpful
to search for the cooling break feature in other hot spots.

\section*{Data availability}
The \textit{Herschel} data utilised in this paper are available from the Herschel science archive.

\section*{Acknowledgements}
\textit{Herschel} is an ESA space observatory with science instruments provided by European-led Principal Investigator consortia and with important participation from NASA.
This work was supported by JSPS KAKENHI Grant Numbers JP21K03635, JP21H01137 and JP18H03721.
We thank Dr. Yukikatsu Terada, Dr. Kosuke Sato and Dr. Satoru Katsuta for their help in writing this paper.

\bibliographystyle{mnras}
\bibliography{bib/reference} 

\bsp	
\label{lastpage}
\end{document}